\documentclass[12 pt]{article}
\usepackage[dvips]{graphicx}
\usepackage{lscape}

\begin{document}
  \title{Cl electrosorption on Ag(100): Lateral interactions and
    electrosorption valency from comparison of Monte Carlo simulations
    with chronocoulometry experiments}
  \author{
    \centerline{ I. Abou Hamad$^{1,2}$, S.J. Mitchell$^{3,4}$,
      Th. Wandlowski$^5$, P.A. Rikvold$^{1,2}$, and G. Brown$^{2,6}$}\\ 
    \centerline{\scriptsize\it $^{1}$Center for Materials Research and 
      Technology 
      and Department of Physics, Florida State University, 
      Tallahassee, FL 32306-4350, USA}\\
    \centerline{\scriptsize\it $^{2}$School of Computational Science,
      Florida State University, Tallahassee, FL 32306-4120, USA}\\
    \centerline{\scriptsize\it $^{3}$Schuit Institute for Catalysis
      and Department of Chemical Engineering,
      Eindhoven University of Technology, 5600 MB Eindhoven, The
      Netherlands}\\ 
    \centerline{\scriptsize\it $^{4}$The Center for Simulational
      Physics and The Department of Physics and Astronomy, 
      The University of Georgia, Athens, GA 30602-2451}\\
    \centerline{\scriptsize\it $^{5}$Institute for Thin Films
      and Interfaces, ISG 3, and Center of Nanoelectronic Systems, cni,
      Research Centre J\"{u}lich, 52425 J\"{u}lich, Germany}\\
    \centerline{\scriptsize\it $^{6}$Center for Computational
      Sciences,
      Oak Ridge
      National Laboratory, Oak Ridge, TN 37831, USA}\\
  }
  \maketitle
  \begin{abstract}
    We present Monte Carlo Simulations using an equilibrium lattice-gas
    model for the electrosorption of Cl on Ag(100)
    single-crystal surfaces. Fitting the simulated isotherms to
    chronocoulometry experiments, we extract parameters such as the
    electrosorption valency $\gamma$ and the next-nearest-neighbor
    lateral interaction energy $\phi_{\rm nnn}$. Both
    coverage-dependent and coverage independent $\gamma$ were previously
    studied, assuming a constant $\phi_{\rm nnn}$~[I.~Abou~Hamad,
      Th.\ Wandlowski, G.~Brown, P.A.~Rikvold, J.\ Electroanal.\
      Chem.\ 554-555 (2003) 211]. Here, a self-consistent, entirely
    electrostatic picture of the lateral interactions with a
    coverage-dependent $\phi_{\rm nnn}$ is developed, and a relationship
    between $\phi_{\rm nnn}$ and $\gamma$ is investigated for Cl on
    Ag(100).
    
  \end{abstract}
  
      
      \section{Introduction}
      \label{sec:I}
      Studies of lateral interactions between adsorbed particles are
      motivated by the need to understand the origin of the wide
      variety of ordered overlayers and phase transitions
      at fractional adsorbate coverage on metal surfaces. These
      interactions have contributions ranging from short-range and
      van der Waals to long-range dipole-dipole, and lattice-mediated
      interactions~\cite{Einstein:96}. Hard-square short-range
      interactions and dipole-dipole long-range interactions are the
      major contributions to the lateral interactions for bromine
      adsorption on Ag(100)~\cite{MitchellSS:01}. In this paper, we
      explore the validity and applicability of such a model for 
      adsorption of chlorine on Ag(100).
      
      Halide electrosorption on single-crystal metal electrode surfaces
      is a good model system for studying the properties of the
      electrode-electrolyte interface in an electrochemical cell. Due to
      its relative simplicity, it can be used to distinguish between
      the various contributions of different parameters according to the
      effect of their inclusion on the overall behavior of the system.
      A mean-field approach is not sufficient, even for the description
      of one of the simplest halide-electrosorption systems Br/Ag(100).
      However, a simple lattice-gas model with constant
      parameters is sufficient to describe its
      equilibrium~\cite{MitchellSS:01,Hamad1}, and
      dynamic~\cite{Hamad2} properties. 
      While the electrosorption of Br on single-crystal Ag(100) from
      aqueous solution has been extensively studied as an example of
      adlayer formation in an electrochemical
      system~\cite{MitchellSS:01,EndoJEAC:99,MitchellJEAC:00,OckoPRL:97,WandJEAC:01},
      less attention has been given to the electrosorption of
      Cl~\cite{Hamad1,MagCR:02,ValetteZPC:78} on Ag(100). A lattice-gas
      model with constant parameters is not sufficient to describe
      Cl/Ag(100), therefore this system can be used to further
      investigate the nature and characteristics of the lateral
      interactions between the adsorbed halide atoms. In particular,
      we here develop a self-consistent picture of variable lattice-gas
      parameters based on the resident charge on the adatoms being
      coverage dependent or electrode-potential dependent (through the
      coverage).

      The rest of this paper is organized as follows. In
      Section~\ref{sec:II} we describe an electrostatic model
      of the adlayer that is used in the simulations, the
      lateral interaction energies, and the Monte Carlo methods
      used. A brief description of the experimental procedure is given
      in Section~\ref{sec:Exp}. The results of fitting the simulations
      to experimental data are detailed in Section~\ref{sec:Results},
      followed by a brief comparison with Br/Ag(100) in
      Section~\ref{sec:compare}. Our conclusions are summarized in
      Section~\ref{sec:Conclusions}.

      \section{Self-consistent electrostatic adlayer model}
      \label{sec:II}
      \subsection{Lattice-gas model}
\begin{figure}
	\includegraphics[angle=0,width=5.5in]{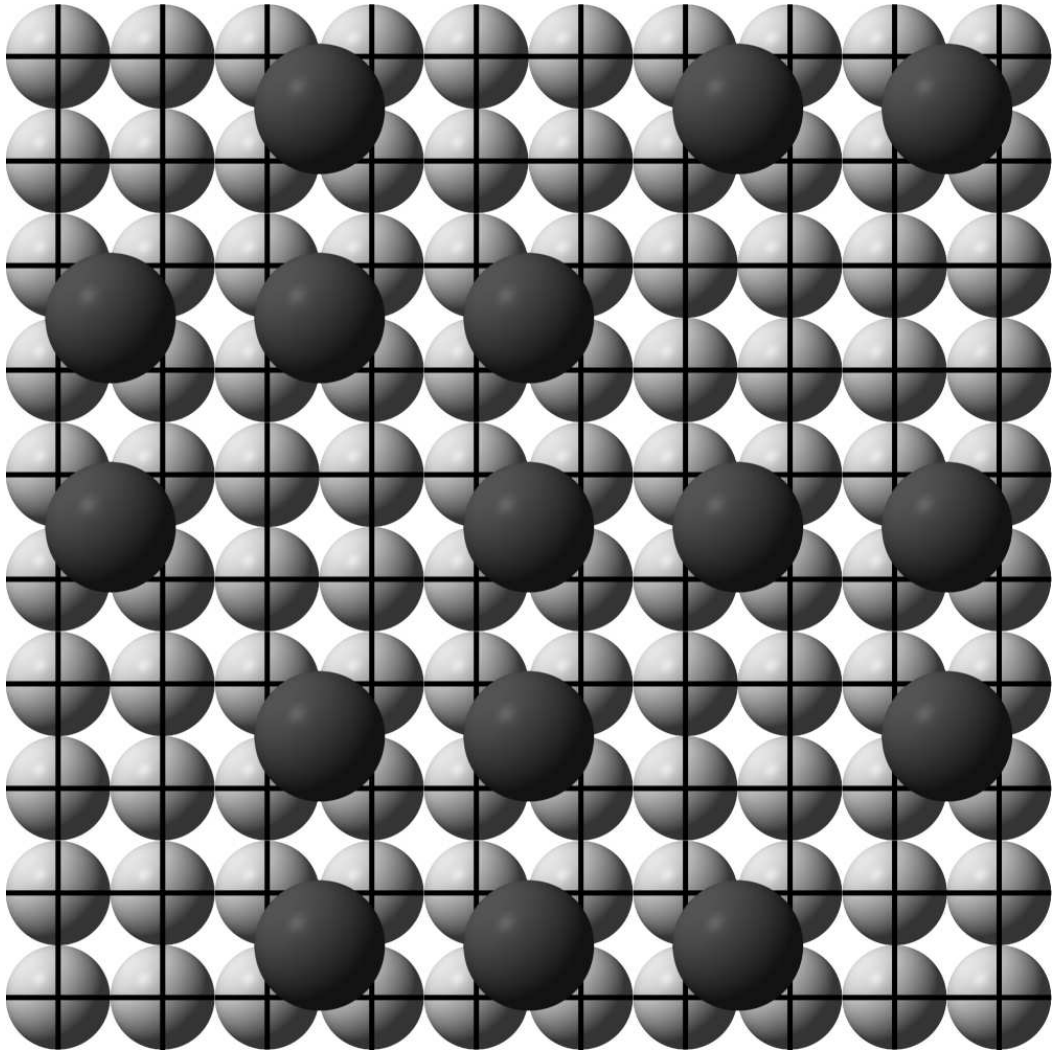}
	\caption[]{Cl (larger, dark gray, spheres) adsorbed at the
	  4-fold hollow sites of the (100) surface of Ag (smaller,
	  lighter gray, spheres). The grid frame corresponds to the
	  lattice of adsorption sites. The figure is drawn
          approximately to scale.
	  \label{fig:model}}
      \end{figure}

      The adsorption of Cl ions occurs at the fourfold hollow sites of
      the Ag(100) surface~\cite{Wand:prep}, which form a square lattice
      as shown in Fig.~\ref{fig:model}. To approximate the equilibrium
      behavior of this system, we use a lattice-gas model, in which
      the lattice sites correspond to the adsorption 
      sites. Mitchell~\textit{et al.}~\cite{MitchellFD:02} used an
      off-lattice model for the Br/Ag(100) system to show that the Br
      adsorbates spend most of the time near the four-fold hollow
      sites of the Ag(100) surface, thus justifying the lattice-gas
      treatment of halide adsorption. To describe the
      energy associated with a configuration of adsorbates on the surface,
      a grand-canonical effective
      Hamiltonian~\cite{MitchellSS:01,MitchellJEAC:00,KoperJEAC:98,KoperEA:98}
      is used,
      \begin{equation}
	{\mathcal{H}} = - \sum_{i<j} \phi_{ij} c_{i} c_{j} - \overline{\mu} 
	\sum_{i=1}^{N}c_{i} \; ,
	\label{eq:H}
      \end{equation}
      where $\sum_{i<j}$ is a sum over all pairs of sites, $ \phi_{ij} $ are
      the lateral interaction energies between particles on the $i$th and
      $j$th lattice sites, measured in meV/pair, $ \overline{ \mu
      } $ is the electrochemical potential, measured in meV/particle, and $N=L^2$
      is the total number of lattice sites. The local
      occupation variable $ c_{i} $ is $1$ if site $i$ is occupied and $0$
      otherwise.
      
      The long-range interactions, $\phi_{ij}$, depend on the distance, $r_{ij}$,
      between ions $i$ and $j$ (measured in Ag(100) lattice spacing units,
      $a=2.889$~{\AA} \cite{OckoPRL:97}) as
      \begin{equation}
	\phi_{ij} = \left\{ \begin{array}{lc}
          - \infty & r_{ij}=1 \\
          \frac{2^{3/2} \phi_{\rm nnn} }{r_{ij}^{3}}
          & r_{ij} \geq \sqrt{2}
	\end{array} \right.,
      \end{equation}
      where the infinite value for $r_{ij}=1$ indicates nearest-neighbor
      exclusion, and negative values of $\phi_{ij}$ denote long-range
      repulsion. The coverage isotherms were
      simulated using a square $L\times L$ lattice with periodic boundary
      conditions to reduce finite-size effects.
      
      The electrochemical potential $\overline{ \mu } $ is related to
      the bulk ionic concentration $C$ and the electrode potential $E$
      (measured in mV). In the dilute-solution approximation, the
      relationship is
      \begin{equation}
	\overline{\mu} = \overline{\mu}_{0} + k_{\mathrm B} T \ln
	\left(\frac{C}{C_{0}}\right)- e \int_{E_0}^{E} \gamma (E'){\rm d} E'\; ,
	\label{eq:mbar} 
      \end{equation}
      where $ \overline{\mu}_{0} $ is an
      arbitrary constant, $C_{0}$ is a reference concentration (here
      taken to be $1~\rm mM$), and $e$ is the elementary charge
      unit~\cite{C}. The reference potential $E_0$ is chosen
      sufficiently negative such that the coverage vanishes at $E_0$
      for all values of $C$ used, and $\overline{\mu}$ has the sign
      convention that $\overline{\mu}>0$ favors adsorption. The
      relationship between $\overline\mu$, $C$, and $E$ is discussed
      further in the Appendix.
      
      \subsection{Lateral interaction energies} 
      When Cl ions adsorb on the surface, a fraction of their charge
      is transfered through the external circuit. This fraction,
      $\gamma e$, is negative and is directly related to the average
      resident charge per ion, $q=-(1+\gamma)e$~\cite{Schmickler:book}.
      This relationship is an approximation and is more valid as the
      potential at the adsorbate approaches the value of the potential
      in the solution. For the current system's ionic strength, this
      condition is only approximately satisfied and may be considered as
      a source of error.

      We have previously shown~\cite{Hamad1} that for Cl/Ag(100), the electrosorption
      valency $\gamma$ depends on the coverage $\theta$, which
      is defined as
      \begin{equation}
	\theta = N^{-1} \displaystyle \sum_{i=1}^{N}{c_{i}}\;.
      \end{equation}
      In order to investigate such a dependence more
      thoroughly, we here propose a model with a coverage-dependent
      next-nearest-neighbor lateral interaction energy $\phi_{\rm nnn}$,
      as well. This is motivated by two assumptions: that $\gamma$ is
      coverage dependent and that the major contribution to $\phi_{\rm nnn}$
      is due to electrostatic dipole-dipole interactions. A 
      simple electrostatic picture of the adlayer, in which an adsorbate's
      resident charge and its image charge form a dipole, suggests a
      relationship between the electrosorption valency and the dipole
      moment. If
      $\gamma$ is coverage dependent ($\gamma=\gamma(\theta)$), then the
      resident charge is also coverage dependent ($q=q(\theta)$), and hence 
      $\phi_{\rm nnn}$, which is proportional to $q^2$, is coverage
      dependent as well ($\phi_{\rm nnn}= \phi_{\rm nnn}(\theta)$).
      
      Assuming for simplicity that $\gamma$ depends linearly on $\theta$,
      \begin{equation}
	\gamma=\gamma_0 + \gamma_1 \theta \;,
      \end{equation}
      the resident charge $q$ becomes
      \begin{equation}
	q=-(1+\gamma_0 + \gamma_1 \theta)e \;,
	\label{eq:q}
      \end{equation}
      and $\phi_{\rm nnn}$ takes the form:
      \begin{equation}
	\phi_{\rm nnn}=A(1+\gamma_0 + \gamma_1 \theta)^2 \;,
	\label{eq:phi}
      \end{equation}
      where $A$ is a prefactor proportional to the square of a
      ``dipole~distance''. This effect is known as
      \textit{depolarization}, and may under extreme conditions
      lead to the possibilty of a first-order phase
      transition~\cite{Roelofs,Koper:98}\footnote{For our model of Cl
      adsorption on Ag(100) with the fitting parameters obtained
      in this paper, the first-order phase transition is not
      observed in the simultions, even for temperatures as
      unrealistically low as  $1$ K. The model does produce a first-order
      transition at room temperature for some parameters that do not fit
      the experimental data.}. If $\gamma$ is coverage
      independent (i.e., $\gamma_1=0$), then $\phi_{\rm nnn}$ becomes
      coverage independent, as well. 
      
      \subsection{Monte Carlo method}
      Equilibrium Monte Carlo (MC) simulations were used to measure
      the equilibrium coverage as a function of $\overline\mu$, which
      was then converted to $E$ using Eq.~(\ref{eq:mbar}). At
      each MC step a lattice
      site, $i$, was chosen randomly, and a change in its occupation
      variable $c_i$ was attempted with a Metropolis acceptance
      probability~\cite{LandauCUP:00}:
      \begin{equation}
	\mathcal{R} = \min \left[ 1,
	  \exp \left(-\frac{\Delta \mathcal{H}}{k_{\rm B}T} \right) \right]\; ,
	\label{eq:P}
      \end{equation}
      where $\Delta \mathcal{H}$ is the energy difference between the
      initial state and the proposed state. The value of $\phi_{\rm nnn}$ was
      updated in the simulations at each MC step, corresponding to the new
      proposed coverage.
      
      The energy difference $\Delta \mathcal{H}$ was calculated
      from Eq.~(\ref{eq:H})
      using two methods, a truncated sum of the contributions 
      of neighboring occupied sites up to five lattice
      spacings away~\cite{MitchellSS:01,Hamad1,MitchellJEAC:00},
      and a mean-field-enhanced truncated sum up to three lattice
      spacings~\cite{Hamad1}. In the latter method, energy 
      contributions from adparticles more than three lattice spacings
      away are calculated using a mean-field estimate as detailed
      in Ref.~\cite{Hamad1}.
      \section{Experimental}
      \label{sec:Exp}
      A detailed description of the experimental procedure is given
      in Ref.~\cite{Hamad1,WandJEAC:01}, and only a brief summary follows here.
      The Ag(100) single-crystal electrodes were chemically etched
      in cyanide solution, rinsed in Milli-Q water, and carefully
      annealed in a hydrogen flame. They were then quickly transfered
      into the electrochemical cell after cooling in a stream of Argon.
      Using the hanging meniscus technique, a platinum wire counter
      electrode, and a saturated calomel electrode as reference electrode,
      the electrochemical measurements were carried out at a temperature of
      ($20 \pm 1$)$^\circ$C.
      
      For the chronocoulometric experiments, the potential was set
      at initial values between $-$1.375 and $-$0.300 V vs SCE until
      adsorption equilibrium was established, and then stepped to the
      final potential of $-$1.400 V, where the chloride is completely
      desorbed from the surface.
      \section{Results}
      \label{sec:Results}
\begin{figure}
	\includegraphics[angle=0,width=5.5in]{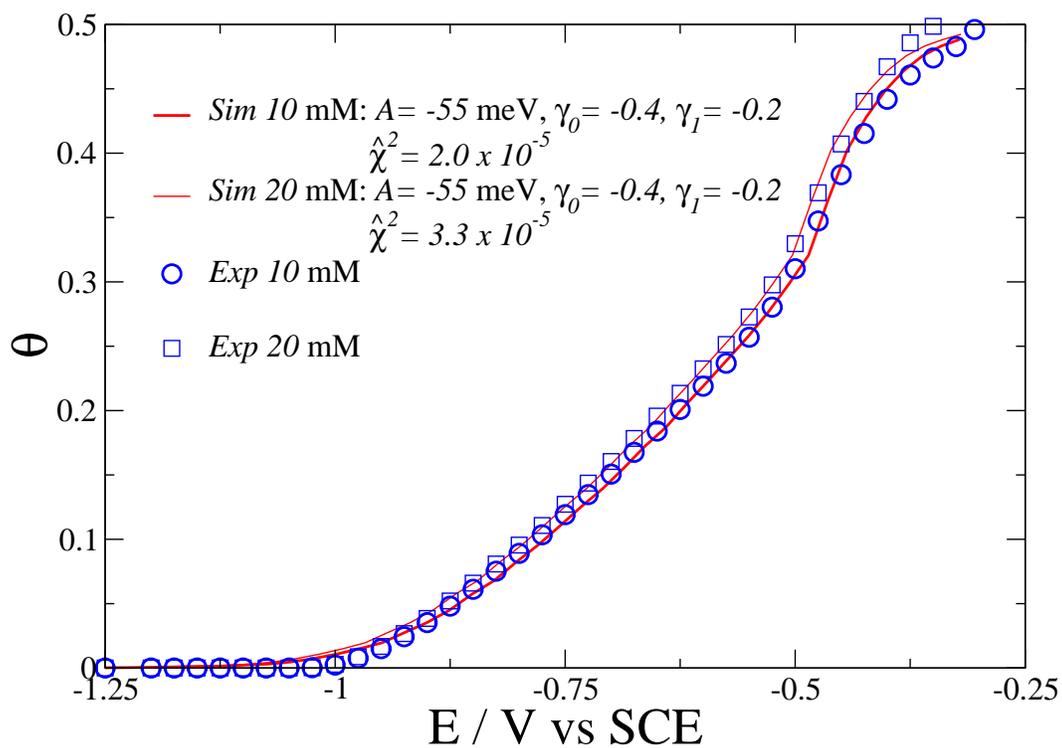}
	\caption{Best fit of mean-field-enhanced simulation
          to experimental 10 mM and 20 mM coverage
          isotherms: $A=-55$ meV, $\gamma_0=-0.4,$ and
	  $\gamma_1=-0.2$ (model~(iii)). $L=32$.
         \label{fig:bestfit}}
      \end{figure}

      The equilibrium MC simulations were performed at a temperature
      of $17^\circ$C for various values of $A,\;\gamma_0$, and
      $\;\gamma_1$.  The value of $A$ was varied from $0$ to
      $-100$ meV in steps of $5$ meV, while $\gamma_0$ and $\gamma_1$
      were varied from $0$ to
      $-0.9$ in steps of $0.1$, resulting in $2000$ different
      simulations. The resulting simulated isotherms were fitted to
      experimental data. This large number of simulations necessitated
      the choice of the relatively small system size with $L=32$. The
      system displays a second-order phase transition from a disordered,
      low-density phase to an ordered $c(2\times2)$
      phase~\cite{MitchellSS:01} at an intermediate value of the
      electrochemical potential. For the steepest part of the isotherm
      (Fig.~\ref{fig:bestfit}), where critical slowing down due to the
      phase transition is important, the system was allowed to
      relax for a longer time to reach equilibrium than for the points
      further away from the phase transition. We checked for
      finite-size effects, and no significant difference was observed
      when comparing the resulting isotherms with isotherms simulated
      using larger system sizes of $L=64, 128$, and $256$. A small
      value of $L$ can be used because $\theta$ is not the order
      parameter corresponding to the $c(2\times2)$ phase, and so its
      fluctuations, which are proportional to
      d$\theta/$d$\overline\mu$, only diverge logarithmically with
      $L$~\cite{MitchellSS:01,Binder:92,Rikvold:85}. In addition,
      better statistics were collected close to the phase transition,
      using longer runs and more sampling than for points far from the
      phase transition, where the fluctuations are smaller.

      Using Eq.~(\ref{eq:mbar}), the simulated isotherms were
      converted from the $\overline\mu$ scale to the $E$ scale and then
      fit to the experimental isotherms. The fitting was done by
      varying the value of $\overline\mu_0$ to minimize
      $\hat{\chi}^2$ for each set of values for $A$, $\gamma_0$,
      and $\gamma_1$. Here,
      \begin{equation}
	\hat{\chi}^2 = \frac{\displaystyle\sum_{\rm exp\;points} (\theta_{\rm
	    exp}-\theta_{\rm sim})^2}{\rm{degrees\;of\;freedom}}
      \end{equation}
      is the least-squares sum per degree of freedom,
      and $\theta_{\rm exp}$ and $\theta_{\rm sim}$ are the
      experimental and simulated coverages, respectively, corresponding
      to the same value of $E$. Linear interpolation was used between
      the simulated data points to calculate $\theta_{\rm sim}$ for
      all values of $E$.
      
      From fits to $10$ mM and $20$ mM experiments, a grid of
      of values of $\hat{\chi}^2$, was
      collected. Three different models were studied. (i) A constant
      $\gamma$ (i.e., also constant $\phi_{\rm nnn}$) model for which
      we only considered the simulations with $\gamma_1=0$. (ii) A
      model in which only the line with
      $\gamma_0=-0.3,\;\gamma_1=-0.3$ was considered. These values for
      $\gamma_0$ and $\gamma_1$ were estimated from values of $\gamma$
      obtained from the experimental data in Ref.~\cite{Hamad1} using
      the method in Ref.~\cite{Lipkowski:book}. (iii) The third model
      studied was the coverage dependent model with
      $\phi_{\rm nnn}=\phi_{\rm nnn}(\gamma(\theta))$, where no constraints on
      the values of any of the parameters were imposed, and the whole
      parameter space was searched for a global minimum.

\begin{landscape}
\begin{figure}
\includegraphics[angle=0,width=6in]{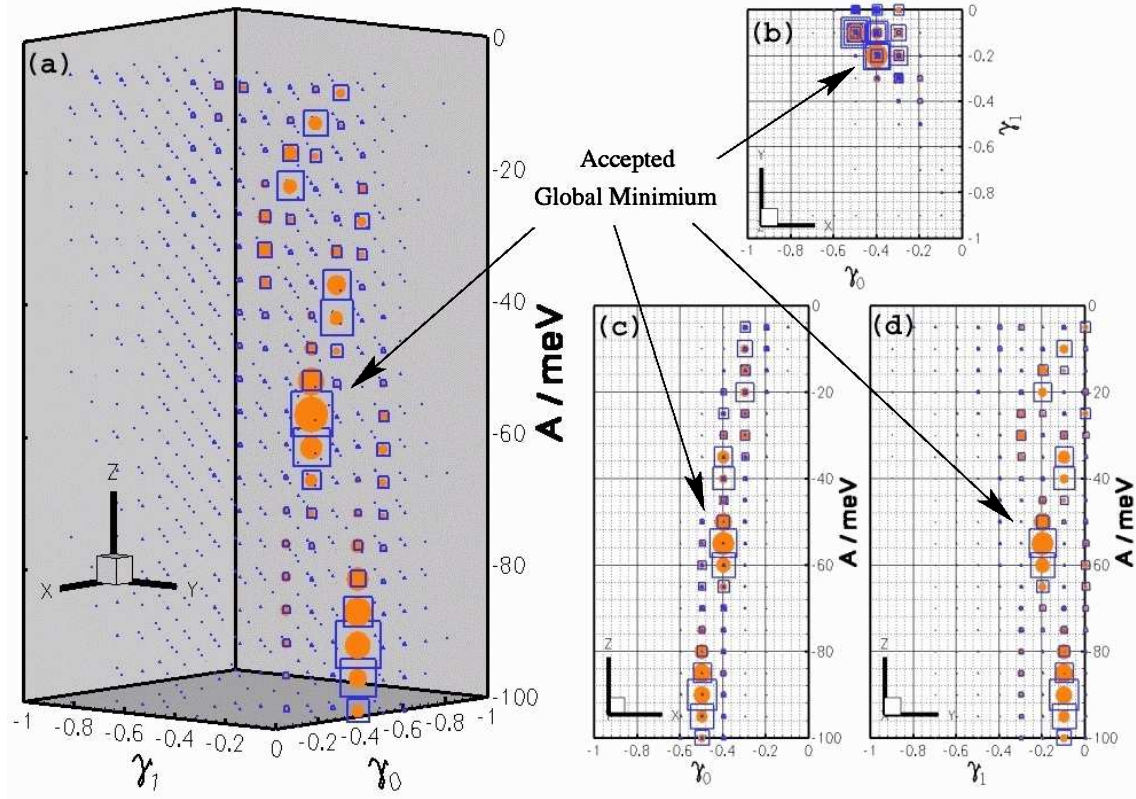}
\caption[]{A plot of $1/\hat{\chi}^2$ in the three-dimensional
parameter space, obtained from fits of mean-field-enhanced simulations
to the 10 mM experimental data (squares) and the 20 mM
experimental data (circles). The diameters of the symbols are
proportional to $1/\hat{\chi}^2$. (a) is a three dimensional view, (b)
is a projection onto the $\gamma_0$, $\gamma_1$ plane,
and (c) and (d) are projections onto the $\gamma_0$, $A$ plane
and the $\gamma_1$, $A$, plane respectively. The parameter space
for the non-mean-field-enhanced method (not shown) is similar.
\label{fig:param}}
\end{figure}
\end{landscape}

      A three-dimensional plot of the parameter space is shown in
      Fig.~\ref{fig:param}(a), where the diameters of the symbols
      are proportional to $1/\hat{\chi}^2$ for fits to $20$~mM~(circles)
      and 10 mM~(squares) of the mean-field-enhanced simulations.
      The parameter space for the non-mean-field-enhanced method
      (not shown) is similar. Figure~\ref{fig:param}(a) shows
      the existence of several local minima and a good overlap between the
      minima for both concentrations. The position of the accepted global
      minimum is indicated by the arrows. Figure~\ref{fig:param}(b) is a 
      projection onto the $\gamma_0$, $\gamma_1$ plane, which shows
      that the minima are concentrated within one region close to the
      $\gamma_1=0$ plane, suggesting a relatively weak dependence
      of $\gamma$ on the coverage. Moreover, the $\hat{\chi}^2$ values in the
      $\gamma_1=0$ plane (model~(i)) and for
      $\gamma_0=\gamma_1=-0.3$~(model~(ii)) 
      are significantly larger than the $\hat{\chi}^2$
      values for the accepted global minimum (model~(iii)).
      Figures~\ref{fig:param}(c) and~\ref{fig:param}(d) show
      that while the magnitude of $\gamma_0$ decreases monotonically 
      with increasing $A$, there is no general trend for
      $\gamma_1$ as a function of
      $A$.

      Due to the existence of several shallow minima of
      $\hat{\chi}^2$, and due to the limited resolution of the grid in
      parameter space, we list in Table~1 the best-fit parameter
      values for each model, along with all fits that have
      $\hat{\chi}^2$ within $10\%$ of the best-fit value.
      One can see in Table~1 that there are two possible sets of parameters
      that fit to the 10~mM data for model~(iii) with mean-field-enhanced
      simulations. To discriminate between these possible fits to the
      10~mM data we check if they are also possible fits to the 20~mM
      experimental data. The first set ($A=-90$ meV) is the overall
      best fit for 10 mM. However, it fits only slightly better than
      the second set ($A=-55$ meV), and it is not a possible fit to the
      20~mM experimental data. The only 
      parameter set ($A=-55$ meV, $\gamma_0=-0.4$, $\gamma_1=-0.2$)
      that is a possible fit to both the 10~mM and 20~mM experimental data,
      is the accepted best fit for model~(iii). Using the same approach
      for the non-mean-field-enhanced method, models (i) and (ii), and concentration,
      results in a unique set of parameter values for each model,
      simulation method, and concentration. These accepted fits
      are summarized in Table~2. Notice in Table~2 that for the accepted
      global minimum of model~(iii), and for both mean-field-enhanced and 
      non-mean-field-enhanced methods, the values of $\gamma_0$ and
      $\gamma_1$ are reasonably close to the ones calculated from
      the experimental data~\cite{Hamad1} using the method of
      Ref.~\cite{Lipkowski:book}. Also note that the fits for 10~mM are
      consistently better than the
      fits to the 20~mM experiments. This might be due to violation of the
      dilute-solution limit at the higher concentration. This may also be
      the reason that the values of $\overline{\mu}_0$, which are expected
      to be the same for both concentrations when fit to a single
      simulation, differ consistently by about $12 \pm 7$ meV between
      the concentrations. The values of $A$ are consistently less negative
      for the longer-ranged lateral interactions of the mean-field-enhanced
      than for the non-mean-field method. This is not surprising.
      
\begin{figure}
	\begin{center}
	  \includegraphics[angle=0,width=5.5in]{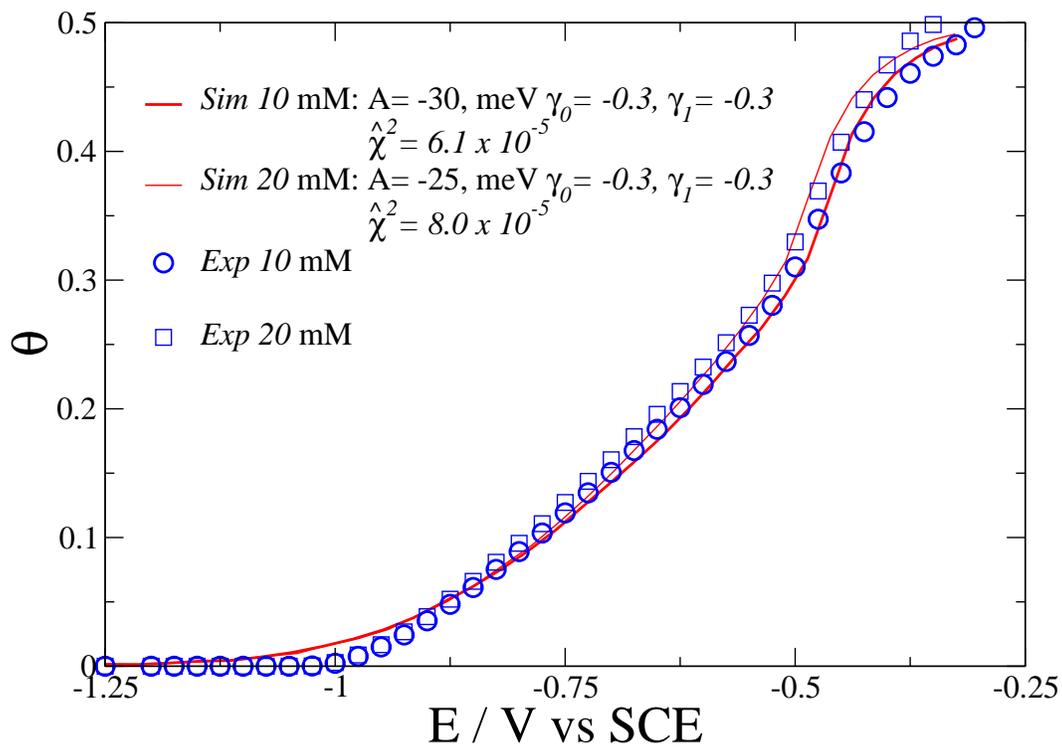}
	  \caption[]{Simulated constrained best fit with the values
	    of $\gamma_0=-0.3$ and $\gamma_1=-0.3$ (model~(ii)) for the 10
	    and 20 mM experimental data. Mean-field-enhanced simulations
            are shown here. $L=32$.
	    \label{fig:expgamma}}
	\end{center}
      \end{figure}

\begin{figure}
	\begin{center}
	  \includegraphics[angle=0,width=5.5in]{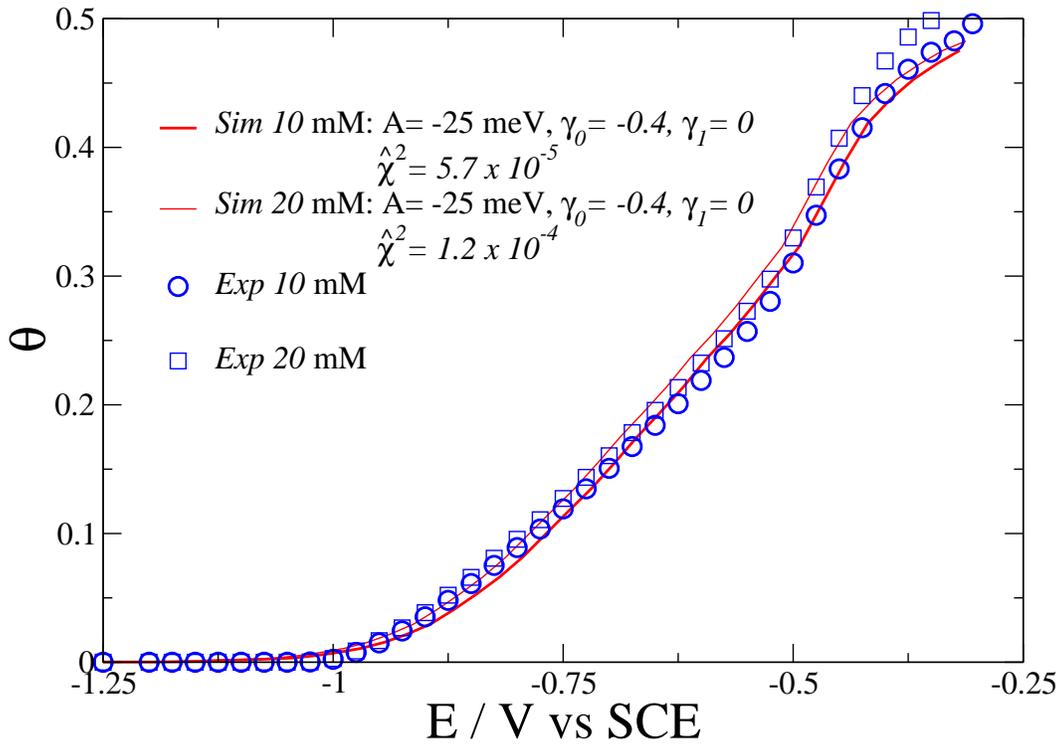}
	  \caption[]{Simulated constrained best fit with a constant
            $\gamma$ (model~(i)) for the 10 and 20 mM experimental
            data. Mean-field-enhanced simulations are shown here. $L=32$.
	    \label{fig:consgamma}}
	\end{center}
      \end{figure}

\begin{figure}
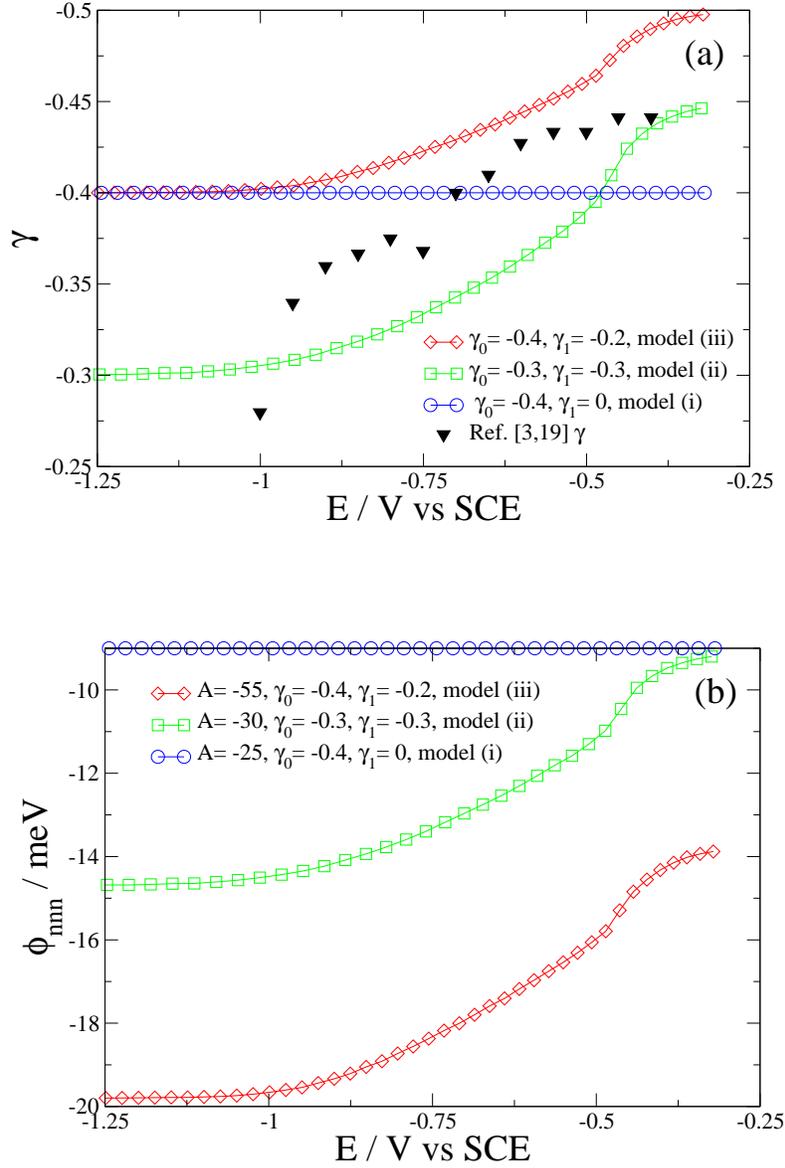

\begin{center}
\includegraphics[angle=0,width=4in]{gamma_vs_E_all}
\mbox{}
\vspace{0.6in}
\mbox{}
\includegraphics[angle=0,width=4in]{PHIofE}
\caption[]{(a) The electrosorption valency $\gamma$ vs $E$ for the
  values of $\gamma_0$ and $\gamma_1$ obtained from the best fits 
  for models (i), (ii), and (iii) to the 10 mM experimental data. The
  values of $\gamma$ obtained in Ref.~\cite{Hamad1} by the method of
  Ref.~\cite{Lipkowski:book} are also shown. (b) Corresponding plots
  of $\phi_{\rm nnn}$ vs $E$ for the three models.\label{fig:gamma_vs_E}}
\end{center}
\end{figure}

      The quality of the fits is better for model~(iii), as can be seen
      from the plots corresponding to each of the
      models~(Figs.~\ref{fig:bestfit},~\ref{fig:expgamma},~and~\ref{fig:consgamma}).
      Models (ii) and (i) fit worse at either the lower-coverage part
      of the isotherm (model~(ii)), see Fig.~\ref{fig:expgamma}, or at
      the upper part of the isotherm (model~(i)), see
      Fig.~\ref{fig:consgamma}. In contrast, the best-fit simulations in
      Fig.~\ref{fig:bestfit} fit
      the experiments well over the whole range of coverages. The plots
      for the non-mean-field-enhanced fits (not shown) are similar. A plot
      of $\gamma$ vs $E$ is shown is Fig.~\ref{fig:gamma_vs_E}(a) for
      the best-fit values of $\gamma_0$ and $\gamma_1$ corresponding
      to the 10 mM experimental data for each of the three models
      considered. Also shown are the values of $\gamma$ obtained in
      Ref.~\cite{Hamad1} by the method of Ref.~\cite{Lipkowski:book}.
      While model~(ii) is expected to fit the
      Ref.~\cite{Hamad1,Lipkowski:book} values of $\gamma$, model~(iii)
      also predicts a mean value of $\gamma$ close to, but slightly more
      negative than, the mean
      of the values of $\gamma$ from Ref.~\cite{Hamad1,Lipkowski:book}.
      The $\phi_{\rm nnn}$ values corresponding to
      the best fits of the three models are shown in
      Fig.~\ref{fig:gamma_vs_E}(b).
      
      Since the fits are sometimes better for the mean-field-enhanced
      method, and sometimes worse, the mean of the sets of parameters
      obtained by including and excluding the mean-field enhancement
      are reported here as our final results:
      $A=-60\;\rm{meV,}\;\gamma_0=-0.4,\;\gamma_1=-0.2$, and $\overline{\mu}_0=-325$ meV.
      In the far-field approximation used here, the lateral
      interaction energy between two parallel dipoles separated by a
      distance $a\sqrt2$ is given by
      \begin{equation}
      \phi_{\rm nnn}=\frac{1}{4\pi\epsilon_0}\frac{p^2}{(a\sqrt2)^3}=A(1+\gamma_0+\gamma_1\theta)^2=A \left( \frac{q}{e}\right)^2\;,
      \end{equation}
      where $p=|q|d$ is the dipole moment, $d$ is an effective
      ``dipolar distance,'' and the second and third equalities result
      from Eqs.~(\ref{eq:phi}) and (\ref{eq:q}), respectively. Consequently, 
      \begin{equation}
      p=\left( \frac{|q|}{e}\right)\sqrt{4\pi\epsilon_0 (a\sqrt2)^3 A}.
      \end{equation}
      For the $A=-60$ meV final result, the dipole moment ranges
      between $p=0.32$~e\AA~for $\theta=0$ and $p=0.26$~e\AA~for
      $\theta=0.5$, and the ``dipolar distance'' $d\approx0.53$~\AA.
      Density Functional Theory calculations of a model without
      water~\cite{DFT} suggest a dipole moment of $0.46$~e\AA,~which
      is only about $40\%$ higher than the fitted value at low coverage.
      From  na\"{\i}ve physical intuition, the dipolar distance
      might be expected to be of the order of the ionic diameter of Cl,
      about $3$\AA. The value of $d$ obtained here involves several
      assumptions. One main assumption used is that the dipole is a
      classical point dipole, which should be quite reasonable as the
      distance between neighboring dipoles is about an order of
      magnitude larger than the obtained value for $d$.
      Moreover, we have assumed that the interaction is a classical
      electrostatic dipole interaction, while there could be quantum
      effects at the shortest length scales. On the other hand, when
      viewed as a charge distribution, the dipole is expected 
      to have a much smaller ``dipolar distance'' than na\"{\i}vely
      expected~\cite{Schmickler:88}. Using the relationship between
      the dipole moment and the Helmholtz capacitance
      $C_{\rm H}$~\cite{Schmickler:book},
      \begin{equation}
      p=|q|d=\frac{e\epsilon_0}{C_{\rm H}}(1+\gamma),
      \end{equation}
      and our values for the dipole moment and $\gamma$,
      we obtain a $C_{\rm H}$ that is in the range of $26$
      to $32$ $\mu$F/cm$^2$. This range is comparable to the reported
      value of $C_{\rm H}$ for Ag(110) in chloride ion solution,
      $C_{\rm H}\approx100\;\mu$F/cm$^2$~\cite{Schmickler:88},
      which would yield $d=\epsilon_0/C_{\rm H}\approx0.1$ \AA.
      This value of $d$ is of the same order of magnitude as our value.
      
      Finally, although models (i) and (ii) are still possible, a
      self-consistent ($\phi_{\rm nnn}(\gamma(\theta))$) entirely
      electrostatic model not only fits better to the experimental data but also,
      with no constraints imposed, predicts values of $\gamma$ that
      are reasonably compatible with those obtained from the
      experimental data using the method of Ref.~\cite{Lipkowski:book}.

      \section{Comparison to Br/Ag(100)}
      \label{sec:compare}
      Br electrosorption on the same substrate Ag(100) displays different
      characteristics. The electrosorption valency and the next-nearest-neighbor
      lateral interaction energy are not only more negative,
      $\gamma=-0.71\pm0.01$ and $\phi_{\rm nnn}=-21\pm2$ meV, but also
      independent of the coverage~\cite{Hamad1}. In other words, while
      mutual depolarization~\cite{Koper:98} is present for Cl/Ag(100),
      this effect is not significant for Br/Ag(100). Simulations
      of Br/Ag(100) using $\gamma=\gamma_0+\gamma_1 \theta$, when
      fit to three experimental data sets, yielded
      $\gamma_1\approx0$~\cite{Hamad1}. Moreover, the value obtained
      for $\gamma$ $(-0.71)$, is consistent with previous results from both
      simulations and experimental analysis~\cite{MitchellSS:01}. Since Cl
      is more electronegative than Br, it is expected to have a less
      negative $\gamma$ or a more negative resident charge, consistent with our
      results. 

      \section{Conclusion}
      \label{sec:Conclusions}
      A discrepancy between the values of the electrosorption valency
      $\gamma$ obtained from fitting simulations to experiments, and
      the value obtained from the analysis~\cite{Lipkowski:book} of
      experimental data was reported in a previous study by
      Abou Hamad \textit{et al.}~\cite{Hamad1}. This discrepancy
      suggested that the long-range interactions are dominated by 
      electrostatic dipole/dipole effects for Cl electrosorption
      on the Ag(100) single crystal surface. 

      In this work, a large set of MC simulations ($2000$ simulations)
      over a grid in parameter
      space was fitted to two sets of chronocoulometry experimental
      data for different concentrations. The existence of local
      minima of $\hat{\chi}^2$
      in parameter space suggests alternative models. To within the
      resolution of our grid and the accuracy of the experimental data,
      we have shown that while other, simpler, models are still
      possible, a purely electrostatic model can be used to describe
      the Cl/Ag(100) system. It also predicts an electrosorption valency
      that is compatible with the value obtained through direct
      experimental data analysis by the method of Ref.~\cite{Lipkowski:book}.
      Additional sets of experimental data, with different concentrations
      around 10 and 20 mM of the adsorbate ion, along with a finer grid in
      parameter space would enable a more decisive determination of the most
      appropriate model and its parameter values.

      \section*{Acknowledgments}

      We thank M.T.M.~Koper and S.~Frank for useful discussions and helpful
      comments. We also thank M.T.M. Koper for bringing
      Refs.~\cite{Roelofs,Koper:98} to our attention. This work was supported
      in part by NSF grant No.~DMR-0240078, and by Florida State 
      University through the School of Computational
      Science and the Center for Materials Research and Technology,
      the Research Centre J\"{u}lich, and the Netherlands Organization for
      Scientific Research (NWO).
 
      \appendix
      \begin{center}
      {\bf Appendix}
      \end{center}
      In this appendix we discuss the relationship between the electrochemical
      potential $\overline \mu$ and the electrode potential $E$ for a coverage
      or field dependent electrosorption valency $\gamma$.

      We define the electrosorption valency as~\cite{Schmickler:book}
      \begin{equation}
       \gamma = -\left(\frac{\partial \sigma_{\rm M}}{\partial \theta}\right)_E\;,
      \end{equation}
      where $\sigma_{\rm M}$ is the charge on the metal, $\theta$ is
      the coverage, and $\mu$ is the \textit{chemical} potential of
      the solution (here taken as the dilute-solution approximation,
      $\mu=k_{\rm B}T\ln (\frac{C}{C_0}) $). Considering an auxiliary
      thermodynamic potential $X$ related to the surface tension
      $\alpha$ as
      \begin{equation}
	X=\alpha + \theta \mu\;,
      \end{equation}
      and using the Lippman equation,
      \begin{equation}
	{\rm d}X = \mu {\rm d}\theta - \sigma_{\rm M} {\rm d}E\;,
      \end{equation}
      we have
      \begin{equation}
	\left(\frac{\partial X}{\partial E}\right)_{\theta} =
	-\sigma_{\rm M}\;{\rm and}\;\left(\frac{\partial X}{\partial
	\theta}\right)_{\sigma_{\rm M}} = \mu\;.
      \end{equation}
      Consequently, one obtains the Maxwell relation,~\cite{Schmickler:book}
      \begin{equation}
	\gamma = -\left(\frac{\partial \sigma_{\rm M}}{\partial
	\theta}\right)_E = \frac{\partial X}{\partial E\partial
	\theta} = \left(\frac{\partial \mu}{\partial E}\right)_{\theta}\;.
      \end{equation}

      The relation, $\gamma = \left(\frac{\partial \mu}{\partial E}\right)_{\theta}$,
      gives the electrosorption
      valency as the change in $\mu$ necessary to keep $\theta$
      constant under a change in $E$. But to keep $\theta$ constant,
      the \textit{electrochemical} potential $\overline\mu$ must
      remain constant, so that
      \begin{equation}
	\gamma = \left(\frac{\partial \mu}{\partial E}\right)_{\theta}
        \equiv \left(\frac{\partial \mu}{\partial E}\right)_{\overline{\mu}}.
      \end{equation}
      
      It is easy to show that Eq.~(\ref{eq:mbar}) satisfies this requirement
      for $\gamma$. If
      \begin{eqnarray*}
      \overline{\mu} = \overline{\mu}_{0} + k_{\mathrm B} T \ln
        \left(\frac{C}{C_{0}}\right)- e \int_{E_0}^{E} \gamma (E'){\rm d} E'\; ,
      \end{eqnarray*}
      then 
      \begin{equation}
	\mu = \overline\mu + e \int_{E_0}^{E} \gamma (E'){\rm d} E' +
	\rm{const.}\; ,
      \end{equation}
      and 
      \begin{equation}
        \left(\frac{\partial \mu}{\partial E}\right)_{\theta}
        = \left(\frac{\partial \mu}{\partial
        E}\right)_{\overline{\mu}}= \gamma (E)\;.
      \end{equation} 
      Generalization to multiple adsorbates is straightforward.

\begin{table}
\caption[]{Fitting parameters for Cl adsorption on a Ag(100) single-crystal
surface, for models (i), (ii), and (iii), with and without
mean-field interactions. Possible fits to within $10\%$ of the
best $\hat{\chi}^2$ for each model are also shown. $A$ and $\overline\mu_0$
are in units of meV.
}
\begin{small}
\begin{tabular}{|c|c|c|c|c|c|c|c|c|c|c|}
\hline
\multicolumn{1}{|c|}{} 
  & \multicolumn{10}{|c|}{Mean-field enhanced} \\
\cline{2-11}      
\multicolumn{1}{|c|}{Model}
  &\multicolumn{5}{|c|}{10 mM}
    &\multicolumn{5}{|c|}{20 mM} \\
\cline{1-11}
  &$A$&$\gamma_0$&$\gamma_1$&$\overline{\mu}_0$&$\hat{\chi}^2\times10^5$
  &$A$&$\gamma_0$&$\gamma_1$&$\overline{\mu}_0$&$\hat{\chi}^2\times10^5$ \\
\cline{2-11}
\multicolumn{1}{|c|}{(i) Const. $\gamma$}
  &$-25$&$-0.4$&$0$&$-330$&$5.664$ &$-60$&$-0.5$&$0$&$-390$&$11.927$ \\
\cline{2-11} 
  &$-5$&$-0.3$&$0$&$-281$&$5.769$&$-55$&$-0.5$&$0$&$-396$&$11.942$ \\
\cline{2-11}  
  & & & & & &$-25$&$-0.4$&$0$&$-340$&$12.079$ \\
\cline{2-11} 
  & & & & & &$-20$&$-0.4$&$0$&$-349$&$12.514$ \\
\hline 
\multicolumn{1}{|c|}{(ii) Ref.~\cite{Hamad1,Lipkowski:book} $\gamma$}
  &$-30$&$-0.3$&$-0.3$&$-266$&$6.068$ &$-25$&$-0.3$&$-0.3$&$-285$&$8.037$ \\
\hline
\multicolumn{1}{|c|}{(iii) $\phi_{\rm nnn}(\gamma(\theta))$}
  &$-90$&$-0.5$&$-0.1$&$-371$&$1.815$ &$-55$&$-0.4$&$-0.2$&$-328$ &$3.340$ \\
\cline{2-11}
  &$-55$&$-0.4$&$-0.2$&$-318$&$1.976$ & & & & & \\
\hline
\hline
\multicolumn{1}{|c|}{} 
    & \multicolumn{10}{|c|}{No mean-field enhancement} \\
\cline{2-11}      
\multicolumn{1}{|c|}{(i) Const. $\gamma$}
  &$-30$&$-0.4$&$0$&$-330$&$5.369$& $-70$&$-0.5$&$0$&$-393$&$11.246$ \\ 
\cline{2-11} 
  & & & & & &$-30$&$-0.4$&$0$&$-341$&$11.330$ \\  
\cline{2-11}  
  & & & & & &$-65$&$-0.5$&$0$&$-398$&$11.350$ \\ 
\cline{2-11} 
  & & & & & &$-25$&$-0.4$&$0$&$-348$&$11.606$ \\ 
\hline 
\multicolumn{1}{|c|}{(ii) Ref.~\cite{Hamad1,Lipkowski:book} $\gamma$}
  &$-35$&$-0.3$&$-0.3$&$-268$&$6.086$ &$-30$&$-0.3$&$-0.3$&$-286$&$7.462$ \\
\hline
\multicolumn{1}{|c|}{(iii) $\phi_{\rm nnn}(\gamma(\theta))$}
  &$-45$&$-0.4$&$-0.1$&$-325$&$2.310$ &$-65$&$-0.4$&$-0.2$&$-330$&$3.112$ \\
\cline{2-11}
  &$-65$&$-0.4$&$-0.2$&$-320$&$2.311$ & & & & & \\
\hline
\end{tabular}
\end{small}
\end{table}

\begin{table}
\caption[]{The best-fit parameters for Cl adsorption on a
Ag(100) single-crystal surface, for models (i), (ii), and (iii),
with and without mean-field interactions. $A$ and
$\overline\mu_0$ are in units of meV.
}
\begin{small}
\begin{tabular}{|c|c|c|c|c|c|c|c|c|c|c|}
\hline
\multicolumn{1}{|c|}{}
  & \multicolumn{10}{|c|}{Mean-field enhanced} \\
\cline{2-11}
\multicolumn{1}{|c|}{Model}
  &\multicolumn{5}{|c|}{10 mM}
    &\multicolumn{5}{|c|}{20 mM} \\
\cline{1-11}
  &$A$&$\gamma_0$&$\gamma_1$&$\overline{\mu}_0$&$\hat{\chi}^2\times10^5$
  &$A$&$\gamma_0$&$\gamma_1$&$\overline{\mu}_0$&$\hat{\chi}^2\times10^5$ \\
\cline{2-11}
\multicolumn{1}{|c|}{(i) Const. $\gamma$}
  &$-25$&$-0.4$&$0$&$-330$&$5.664$ &$-25$&$-0.4$&$0$&$-340$&$12.079$ \\
\cline{1-11}
\multicolumn{1}{|c|}{(ii) Ref.~\cite{Hamad1,Lipkowski:book} $\gamma$}
  &$-30$&$-0.3$&$-0.3$&$-266$&$6.068$ &$-25$&$-0.3$&$-0.3$&$-285$&$8.037$ \\
\hline
\multicolumn{1}{|c|}{(iii) $\phi_{\rm nnn}(\gamma(\theta))$}
  &$-55$&$-0.4$&$-0.2$&$-318$&$1.976$ &$-55$&$-0.4$&$-0.2$&$-328$ &$3.340$ \\
\hline
\hline
\multicolumn{1}{|c|}{}
    & \multicolumn{10}{|c|}{No mean-field enhancement} \\
\cline{2-11}
\multicolumn{1}{|c|}{(i) Const. $\gamma$}
  &$-30$&$-0.4$&$0$&$-330$&$5.369$& $-30$&$-0.4$&$0$&$-341$&$11.330$ \\
\cline{1-11}
\multicolumn{1}{|c|}{(ii) Ref.~\cite{Hamad1,Lipkowski:book} $\gamma$}
  &$-35$&$-0.3$&$-0.3$&$-268$&$6.086$ &$-30$&$-0.3$&$-0.3$&$-286$&$7.462$ \\
\hline
\multicolumn{1}{|c|}{(iii) $\phi_{\rm nnn}(\gamma(\theta))$}
  &$-65$&$-0.4$&$-0.2$&$-325$&$2.310$ &$-65$&$-0.4$&$-0.2$&$-330$&$3.112$ \\
\hline
\end{tabular}
\end{small}
\end{table}

\end{document}